\documentclass[preprint, 12pt, 3p,
authoryear]{elsarticle} %review=doublespace preprint=single 5p=2 column
\usepackage[hyphens]{url}

\usepackage{graphicx}
\usepackage[T1]{fontenc}
\usepackage{lmodern}
\usepackage{amssymb,amsmath}
\usepackage{lineno} % add
\usepackage{ifxetex,ifluatex}
\usepackage{fixltx2e} % provides \textsubscript
\IfFileExists{upquote.sty}{\usepackage{upquote}}{}
\ifnum 0\ifxetex 1\fi\ifluatex 1\fi=0 % if pdftex
  \usepackage[utf8]{inputenc}
\else % if luatex or xelatex
  \usepackage{fontspec}
  \ifxetex
    \usepackage{xltxtra,xunicode}
  \fi
  \defaultfontfeatures{Mapping=tex-text,Scale=MatchLowercase}
  
\fi
\IfFileExists{microtype.sty}{\usepackage{microtype}}{}
\usepackage[]{natbib}
\ifxetex
  \usepackage[setpagesize=false, % page size defined by xetex
              unicode=false, % unicode breaks when used with xetex
              xetex]{hyperref}
\else
  \usepackage[unicode=true]{hyperref}
\fi
\hypersetup{breaklinks=true,
            bookmarks=true,
            pdfauthor={},
            pdftitle={A Design Concept of Forecasting Software for NormalizedVector Autoregressions with Fat Tails and Stochastic Volatility},
            colorlinks=false,
            urlcolor=blue,
            linkcolor=magenta,
            pdfborder={0 0 0}}

\usepackage[utf8]{inputenc} \usepackage{amsmath} \usepackage{amsthm} \usepackage{amssymb} \usepackage{natbib} \usepackage{booktabs} \usepackage{multirow} \usepackage{setspace}

\emergencystretch=\maxdimen \hyphenpenalty=10000 \hbadness=10000

\begin{document}

\begin{frontmatter}

  \title{A Design Concept of Forecasting Software for
Normalized\linebreak Vector Autoregressions with Fat Tails and
Stochastic Volatility}
    \author[guangdong]{Fei Shang%
  }
  
    \author[unimelb]{Xiaolei Wang%
  }
  
    \author[unimelb]{Tomasz Woźniak}
      \address[guangdong]{School of Economics and Trade, Guangdong University of
Foreign Studies, Guangzhou, 510006, China}
    \address[unimelb]{Department of Economics, University of Melbourne, 111 Barry Street, Carlton, 3053, VIC, Australia}

  \begin{abstract}
  We present a suite of \textbf{R} packages for macroeconomic
  forecasting that leverages advanced Bayesian, structural,
  multivariate, dynamic, hierarchical, non-linear, and non-Gaussian
  models. The suite enables both structural and predictive analyses, and
  is adapted to time series data across various types, dimensions, and
  sampling frequencies. Each additional feature increases computational
  complexity. To address this challenge, our software design
  incorporates a carefully curated selection of models, efficient
  algorithms implemented in \textbf{C++}, advanced econometric and
  numerical methods, robust handling of complex input and output
  objects, and standardised workflows. This approach combines the
  computational efficiency of \textbf{C++} with the convenience of
  working with data in \textbf{R}. We demonstrate that our packages
  facilitate original research contributions in forecasting, as
  illustrated by our example in which vector autoregressions with
  non-centred stochastic volatility enhance density and point
  predictions relative to models with centred stochastic volatility.\\
  \end{abstract}

   \bigskip 
  
 \end{frontmatter}

\section{Introduction}

\noindent Macroeconomic forecasting is crucial for policy-makers,
businesses, and researchers to make informed decisions in an
ever-changing economic landscape. These decisions affect crucial
economic aggregates and other indicators determining well-being at all
future horizons \citep[see][]{rtahhps2022}. The gravity of such
prediction-driven economic planing, institutional design requiring
interpretability, and complex properties of macroeconomic and financial
time series lead to a burgeoning field of enquiry focusing on scalable
dynamic system modelling \citep[see][]{dls1984}. Additionally, the
global financial crisis and the COVID-19 pandemic highlighted the
importance of robust forecasting models that can adapt to unprecedented
shocks and volatility in economic data \citep[see][]{glp2022}.

Our forecasting software design accommodates these requirements ensuring
transparency and reproducibility of predictive analyses. However, its
practical usability faces even more challenges arising from the
numerical complexity of handling Bayesian hierarchical and non-linear
modelling. Moreover, point and density forecasting handles forecasting
of measurements and latent dynamic processes jointly, resulting in high
dimensionality of numerical integration.

In this context, we present a suite of \textbf{R} packages designed for
forecasting using Vector Autoregressions (VARs) with Stochastic
Volatility (SV) and Student-t errors, which are particularly suited for
handling complex non-linear dynamics in time series data. The suite
includes three packages: \textbf{bsvars} by
\cite{bsvars,bsvars_vignette} implementing the heteroskedastic
structural VARs by \cite{lsuw2026} and \cite{sw2026}, package
\textbf{bsvarSIGNs} by \cite{bsvarSIGNs_vignette,bsvarSIGNs} focusing on
the hierarchical VARs for post-COVID forecasting by \cite{glp2015} and
\cite{glp2022}, and package \textbf{bvars} by \cite{bvars} providing the
flexible VAR model by \cite{chan2020large}. The focus is on normalised
models with well-specified covariance matrix of the predictive density,
facilitating scalable robust predictions.

In the current paper, we highlight the bespoke design of the three
open-source packages. Together, they address the challenges through the
following features:

\begin{itemize}\setstretch{1.5}
\item deliberate selection of modelling approaches representing commonly used specifications, 
\item a range of forecasting techniques accommodating best practices in the field, 
\item fast computations by relying on algorithms written in \textbf{C++},
\item supported by adaptation of frontier econometric and numerical methods,
\item handling multi-layered inputs and outputs by object-oriented programming, and
\item coherent forecasting workflows enabling customisation and transparency.
\end{itemize}

Taken together these packages offer an unmatched versatility in dealing
with data and offering ample opportunities for generating original
predictive analyses. We illustrate these opportunities by an example of
forecasting a 10-variable monetary-fiscal system to document the merit
of presented approaches. In particular, the VARs with non-centred SV
proposed by \cite{lsuw2026} and in the current manuscript perform best
in terms of point and density forecasting 1-quarter and 1-year ahead.

\section{A Suite of Predictive Models}\label{sec:models}

\noindent The packages include a selection of state-of-the-art models
that enable both structural and predictive analyses across a wide range
of specifications. This suite of models is created in accordance with
recent literature and established best practices. Specifically, it
features non-normal, heteroskedastic, structural and reduced-form VARs
with hierarchical priors. These characteristics are incorporated based
on substantial evidence supporting their effectiveness in improving
forecasting performance, including that provided by \cite{dls1984} and
\cite{glp2015} for the Minnesota prior and hierarchical prior shrinkage,
by \cite{Clark2011} for conditional heteroskedasticity, by
\cite{chiu2017} and \cite{chan2020large} for non-normal errors, and by
\cite{cky2024} for order-invariance. Also normalisation of predictive
density covariance enhances the models' capacity to handle diverse
datasets and increases the numerical stability of estimation and
forecasting. The packages provide both general models and nested
specifications, as detailed in \cite{bsvars_vignette} and
\cite{bsvarSIGNs_vignette}.

All the models in our packages share the VAR equation with lag order
\(p\) specified for the \(N\)-vector of dependent variables
\(\mathbf{y}_t\) at time \(t\) as follows: \begin{align}
\mathbf{y}_t = \mathbf{A}_1 \mathbf{y}_{t-1} + \ldots + \mathbf{A}_p \mathbf{y}_{t-p} + \mathbf{A}_d \mathbf{d}_{t} + \boldsymbol{\epsilon}_t,\label{eq:var}
\end{align} where \(\mathbf{A}_i\) is either an \(N\times N\)
autoregressive matrix for \(i=1,\ldots,p\) or an \(N\times D\) matrix of
coefficients on exogenous terms collected in \(D\)-vector
\(\mathbf{d}_{t}\) for \(i=d\), and \(\boldsymbol{\epsilon}_t\) is an
\(N\)-vector of error terms. Define an \(N\times (Np+D)\) matrix
\(\mathbf{A} = \begin{bmatrix}\mathbf{A}_1&\dots&\mathbf{A}_p&\mathbf{A}_d\end{bmatrix}\),
and an \((Np+D)\)-vector
\(\mathbf{x}_t = \begin{bmatrix}\mathbf{y}_{t-1}'&\dots&\mathbf{y}_{t-p}'&\mathbf{d}_{t}'\end{bmatrix}'\),
and rewrite equation \eqref{eq:var} as
\(\mathbf{y}_t = \mathbf{A}\mathbf{x}_t + \boldsymbol{\epsilon}_t\).

Furthermore, both of our structural and reduced-form models share an
error term that implies the predictive density's conditional normality
given past information, parameters, and latent processes driving
non-normality and heteroskedasticity: \begin{align}
\text{reduced-form: } && \mathbf{y}_t \mid \mathbf{x}_{t}, \mathbf{A}, \boldsymbol\Sigma, \sigma_t^2, \lambda_t &\sim \mathcal{N}_N\left(\mathbf{A}\mathbf{x}_t, \sigma_t^2\lambda_t\boldsymbol\Sigma\right),\label{eq:rf}\\
\text{structural-form: } && \mathbf{y}_t \mid \mathbf{x}_{t}, \mathbf{A}, \mathbf{B}, \boldsymbol\sigma_t^2, \boldsymbol\lambda_t &\sim \mathcal{N}_N\left(\mathbf{A}\mathbf{x}_t, \mathbf{B}^{-1}\text{diag}\left(\boldsymbol\sigma_t^2\right)\text{diag}\left(\boldsymbol\lambda_t\right)\mathbf{B}^{-1\prime}\right).\label{eq:sf}
\end{align} The reduced-form model in expression \eqref{eq:rf} has its
covariance specified by a scalar common volatility component
\(\sigma_t^2\), a scalar independent latent process \(\lambda_t\), and a
constant \(N\times N\) symmetric, positive-definite, covariance matrix
\(\boldsymbol\Sigma\), following the models proposed by
\cite{carriero2016common} and \cite{chan2020large}. The structural-form
model in expression \eqref{eq:sf} has its covariance specified by a
matrix product involving equation-specific SV components
\(\boldsymbol\sigma_t^2 = \begin{bmatrix}\sigma_{1.t}^2&\dots&\sigma_{N.t}^2\end{bmatrix}'\)
and independent processes
\(\boldsymbol\lambda_t = \begin{bmatrix}\lambda_{1.t}&\dots&\lambda_{N.t}\end{bmatrix}'\),
and an \(N\times N\) non-singular structural matrix \(\mathbf{B}\),
following \cite{lsuw2026}.

The SV components are implemented in two versions, centred and
non-centred. The former was adapted to order-invariant structural VARs
by \cite{cky2024} and to reduced-form VARs by \cite{carriero2016common},
while the latter was adapted to order-invariant structural VARs by
\cite{lsuw2026}. In the current paper, we show the benefits of
forecasting using non-centred SV in both structural and reduced-form
models. To simplify the exposition, the SV models are presented for a
common SV component \(\sigma_t^2\) and apply to each of
\(\sigma_{n.t}^2\): \begin{align}
\text{centred SV: } & \sigma_t^2 = \exp(h_t), \quad h_t = \rho h_{t-1} + \eta_t, \quad \eta_t \sim \mathcal{N}(0, \omega^2),\label{eq:csv}\\
\text{non-centred SV: } & \sigma_t^2 = \exp(\omega h_t), \quad h_t = \rho h_{t-1} + \eta_t, \quad \eta_t \sim \mathcal{N}(0, 1),\label{eq:ncsv}
\end{align} where \(h_t\) is the log-volatility, \(\omega\) a standard
deviation of log-conditional variances, and an autoregressive parameter
\(\rho\). The difference between the two models is in the prior
distribution for \(\omega^2\), that is, inverse gamma in the centred,
and gamma in the non-centred representation.

While the centred SV has been used commonly in macroeconometric
applications, \cite{lsuw2026} argues that the non-centred SV is more
suitable for structural models as it ensures normalisation around the
homoskedastic case, \(\sigma_t^2 = 1\), by the appropriate use of prior
distributions. They argue that normalisation of the system defined by a
multiplicative structure for covariance in \eqref{eq:sf}, is essential
as otherwise the volatility \(\sigma_t^2\) and the structural matrix
\(\mathbf{B}\) are identified up to a normalising constant.

We extend this argument to reduced-form models with a common SV in
\eqref{eq:rf}, and complement it with novel means of normalising the
process \(\lambda_t\). This independent process is responsible for
modelling Student's t predictive density, based on the conditionally
normal specification, and the inverse gamma prior for \(\lambda_t\) with
degrees of freedom \(\nu\) and scale \(\nu - 2\): \begin{align}
\lambda_t\mid\nu \sim \mathcal{IG}2\left(\nu - 2, \nu\right),\text{ for } t\in\{1,\dots,T\}.\label{eq:t}
\end{align} The restriction on the prior scale in \eqref{eq:t}, relating
it to \(\nu\), was proposed by \cite{chan2020large} to normalise the
prior expectation of \(\lambda_t\) to
1.\footnote{The expectation of a random variable following $\mathcal{IG}2\left(s, \nu\right)$ equals $\frac{s}{\nu-2}$ for $\nu>2$ \citep[][]{blr1999}.}
Then, integrating out \(\lambda_t\) from the distributions in
\eqref{eq:rf} and \eqref{eq:sf} results in the marginal Student's t
predictive distribution \citep[see][]{blr1999}. Specifically, a joint
distribution with \(\nu\) degrees of freedom results from \eqref{eq:rf},
and marginal equation-specific distributions with corresponding degrees
of freedom, \(\begin{bmatrix}\nu_1&\dots&\nu_N\end{bmatrix}\), result
from \eqref{eq:sf}. The integration is performed numerically during the
estimation, following the Gibbs sampler, and thus, \(\lambda_t\) is
estimated as part of the procedure. A normal model is nested in this
specification by restricting \(\lambda_t\) to 1 strictly, in which case
neither \(\lambda_t\) nor \(\nu\) are estimated.

However, the prior scale restriction \eqref{eq:t} is insufficient to
normalise the posterior output in systems such as those in \eqref{eq:rf}
and \eqref{eq:sf}, which leads to severe numerical instability.
Therefore, we also restrict the scale of the inverse gamma full
conditional posterior distribution of \(\lambda_t\) given by
\(\mathcal{IG}2(\nu - 2 + u_t^2, \nu +1)\), to ensure that the posterior
expectation of \(\lambda_t\) is equal to 1. It is implemented by
imposing the restriction \(\mathbb{E}[u_{n.t}^2]=1\) in our code, where
\(u_{n.t}\) is an element of
\(\text{diag}\left(\boldsymbol\sigma_t^{-1}\right)\mathbf{B}\boldsymbol{\epsilon}_t\)
in structural models, and
\(\text{chol}\left(\boldsymbol\Sigma^{-1}\right)\boldsymbol{\epsilon}_t/\sigma_t\)
in reduced-form models. In effect, our packages implement effective
approaches to normalising VAR models, resulting in numerically stable
and scalable algorithms suitable for a variety of data sets.

Our packages implement Bayesian forecasting based on the predictive
density. It is defined as the conditional distribution of the future
values \(\mathbf{y}_{T+1}\) given sample data \(\mathbf{Y}_{T}\), with
the forecast origin \(T\). We present it in a simplified form for
one-period-ahead forecast as: \begin{multline}
p\left(\mathbf{y}_{T+1}\mid \mathbf{Y}_{T} \right) = \int p\left(\mathbf{y}_{T+1}\mid \mathbf{Y}_{T}, \boldsymbol\theta, \sigma_{T+1}^2, \lambda_{T+1}\right)\\
\times p\left(\sigma_{T+1}^2\mid \mathbf{Y}_{T}, \boldsymbol\theta\right)
p\left(\lambda_{T+1}\mid \mathbf{Y}_{T}, \boldsymbol\theta\right)
p\left(\boldsymbol\theta \mid \mathbf{Y}_{T}\right) d\left(\boldsymbol\theta, \sigma_{T+1}^2, \lambda_{T+1}\right),\label{eq:fore}
\end{multline} where \(\boldsymbol\theta\) collects parameters of the
model and in-sample latent processes. The right-hand side of
\eqref{eq:fore} is constructed using the the conditional predictive
density,
\(p\left(\mathbf{y}_{T+1}\mid \mathbf{Y}_{T}, \boldsymbol\theta, \sigma_{T+1}^2, \lambda_{T+1}\right)\),
specified by \eqref{eq:rf} or \eqref{eq:sf}, volatility predictive
density,
\(p\left(\sigma_{T+1}^2\mid \mathbf{Y}_{T}, \boldsymbol\theta\right)\),
defined by \eqref{eq:csv} or \eqref{eq:ncsv}, and the predictive density
of the independent process,
\(p\left(\lambda_{T+1}\mid \mathbf{Y}_{T}, \boldsymbol\theta\right)\),
as in \eqref{eq:t}. Finally,
\(p\left(\boldsymbol\theta \mid \mathbf{Y}_{T}\right)\) is the posterior
distribution of parameters given data.

The integral is computed numerically to obtain \(S\) draws from the
predictive density \(\left\{\mathbf{y}_{T+1}^{(s)}\right\}_{s=1}^{S}\).
This procedure involves obtaining \(S\) draws from the posterior
distribution, \(\left\{\boldsymbol\theta^{(s)}\right\}_{s=1}^{S}\), and
the predictive distributions of the latent processes,
\(\left\{\sigma_{T+1}^{2(s)},\lambda_{T+1}^{(s)}\right\}_{s=1}^{S}\).
Finally, for each \(s\), \(\mathbf{y}_{T+1}^{(s)}\) is sampled from
\(p\left(\mathbf{y}_{T+1}\mid \mathbf{Y}_{T}, \boldsymbol\theta^{(s)}, \sigma_{T+1}^{2(s)}, \lambda_{T+1}^{(s)}\right)\).

All our packages implement forecasting in a unified manner and report
the draw from the predictive density as well as its mean and covariance
for each \(s\). The draws can be used to report forecast summaries,
whereas the mean and covariance can be used to compute density
forecasting performance measures.

\section{Forecasting Software Features}\label{sec:software}

\noindent Our packages implement Bayesian, structural, multivariate,
dynamic, hierarchical, non-linear, and non-normal models that are
specifically designed for robust forecasting, adaptability, and
scalability. However, each listed feature comes at the cost of increased
computational complexity. Our design addresses this challenge by relying
on algorithms written in \textbf{C++}, implementing frontier numerical
methods, handling complex input and output objects, and enabling
adjustable forecasting workflows. Therefore, we combine the best of both
approaches: fast algorithms written in \textbf{C++} and the convenience
of data analysis in~\textbf{R}, which has a decisive role in making
these methods available to practitioners.

\subsection{Reliance on Algorithms Written in C++}\label{ssec:cpp}

\noindent In our packages, routines for estimation, forecasting, and
processing of the rich estimation output, such as computing forecast
error variance decomposition, are implemented using \textbf{C++} code.
This task is facilitated by the \textbf{Rcpp} package by
\cite{eddelbuettel2011rcpp}, which automates compilation and linking of
the \textbf{C++} code and ensures object compatibility with \textbf{R}.
We rely heavily on linear algebra and pseudo-random number generators,
which are facilitated using the package \textbf{RcppArmadillo} by
\cite{eddelbuettel_rcpparmadillo2014}, a collection of headers linking
to the \textbf{C++} library \textbf{armadillo} by
\cite{sanderson2016armadillo}.

Both algebraic operations and random number generation are performed
much faster with \textbf{RcppArmadillo} than using many alternative
solutions we tested. For instance, obtaining 1000 random draws from a
standard normal distribution using \textbf{RcppArmadillo},
\texttt{arma::vec\ out(1000,\ arma::fill::randn);}, is 33 per cent
faster on average than \textbf{Rcpp} implementation,
\texttt{Rcpp::NumericVector\ out\ =\ Rcpp::rnorm(1000,\ 0,\ 1);}, with
even larger differential for other solutions. We also use \textbf{C++}
functions for linear algebra on tri-diagonal matrices from package
\textbf{stochvol} by \cite{factorstochvol}, and specialised samplers
from truncated normal and generalised inverse Gaussian distributions
using packages \textbf{RcppTN} by \cite{RcppTN} and \textbf{GIGrvg} by
\cite{GIGrvg}, respectively. All of these developments make the
algorithms computationally fast. Still, estimation of our models is
demanding and may take a while, which we address by providing a progress
bar implemented with the \textbf{RcppProgress} package by
\cite{RcppProgress}.

As an additional feature contributing to open source forecasting
community, our packages export both: essential \textbf{R} functions and
all \textbf{C++} functions using the \textbf{Rcpp} attribute,
\texttt{//\ {[}{[}Rcpp::interfaces(cpp){]}{]}} \citep[see][]{aef2026},
and providing dynamic libraries \textbf{bsvars}, \textbf{bsvarSIGNs},
and \textbf{bvars}. Therefore, developers of \textbf{R} packages can use
our functions directly in their \textbf{C++} code bypassing the imports
of \textbf{R} functions.

\subsection{Numerical Methods}\label{ssec:numeric}

\noindent A number of frontier numerical methods tailored to the
particular model specifications were implemented in our Bayesian
estimation procedures, thereby reducing computational complexity. For
instance, our empirical example applies VARs to 10 quarterly variables,
requiring an intercept and 4 lags and resulting in 410 autoregressive
parameters. Sampling these parameters efficiently is crucial for the
overall performance of the estimation procedure. In reduced-form models
from packages \textbf{bsvarSIGNs} and \textbf{bvars} these parameters
are sampled from a matrix-variate normal full conditional posterior
distribution, the covariance matrix of which is defined by the Kronecker
product of two smaller matrices, greatly reducing the burden of
inverting this matrix \citep[see][]{chan2020large}. Further
computational gains are possible in the structural models in package
\textbf{bsvars}, where these parameters are sampled equation-by-equation
as in \cite{cccm2022}.

Similarly, \textbf{bsvars} implements the row-by-row Gibbs sampler for
the structural matrix proposed by \cite{wz2003}, exhibiting efficient
mixing and fast convergence. Our \textbf{C++} implementation of this
sampler was over 60 times faster than equivalent \textbf{R} code we had
optimised over years, for an empirically relevant structural matrix of
dimensions \(8\times8\). Furthermore, the models in package
\textbf{bsvarSIGNs} apply a feasible decomposition of the joint
posterior distribution into a marginal density for hyper-parameters and
a conditional one for autoregressive parameters and the covariance
matrix, as in \cite{glp2015}. This decomposition results in a
Metropolis-Hastings algorithm for a low-dimensional distribution over
hyper-parameters and an independent sampler for the remaining parameters
parallelised using function \texttt{mclapply} and \texttt{parLapply}
from package \textbf{parallel}.

Optimal approaches to estimating parameters also drive the allocation of
models into the packages. The guiding principle is to gather models
implying similar numerical procedures. The package \textbf{bsvars}
includes structural models that rely on row-by-row estimation of the
parameters, whereas \textbf{bvars} includes reduced-form models. In
these two packages, estimation is performed using Gibbs sampler implying
serial computations. The \textbf{bsvarSIGNs} package focuses on models
relying on independent sampler facilitating parallelisation.

Estimation of conditional heteroskedasticity poses further challenges,
as the dimension of the latent processes driving volatility equals to
the time dimension. Both packages \textbf{bsvars} and \textbf{bvars}
apply the same techniques for the estimation of SV. They rely on the
transformation of the non-linear non-normal state-space model into a
much faster-to-estimate linear, conditionally Gaussian representation
proposed by \cite{ocsn2007}, where the conditional normality is achieved
by approximating the log-\(\chi_1^2\) distribution by a mixture of 10
normal components with fixed parameters. Our common SV models rely on
similar normal mixture approximation of the log-\(\chi_N^2\)
distribution, where \(N\) is the number of variables, estimated using a
fast \textbf{C++} implementation by \cite{sc2017}. This approach to SV
estimation requires sampling from a multivariate normal distribution of
dimension equal to the sample size, which is computationally expensive.
We take advantage of the fact that the covariance matrix of this
distribution is specified by a banded precision matrix as in
\cite{chan2009a}. An appropriate implementation of the code from the
package \textbf{stochvol} facilitates over 10-fold computational gains
relative to less efficient implementations as documented by
\cite{w2021}. Finally, to grant fast convergence of SV estimation,
allowing users to reduce the length of the Markov Chain Monte Carlo
(MCMC), we implement the ancillarity-sufficiency interweaving strategy
by \cite{kfs2014}.

\subsection{Handling Inputs and Outputs}\label{ssec:r6}

\noindent Providing software packages for Bayesian forecasting requires
handling complicated input and output objects. We code these structures
by applying object-oriented programming using the \textbf{R6} package by
\cite{R6} to combine the simplicity of initiating the forecasting
workflows with ample possibilities for customisation. Consider an
example in which the SVAR-SV model from package \textbf{bsvars} is
applied to fiscal policy as in \cite{lsuw2026}.

Begin by investigating the model specification that includes fixed
hyper-parameters of prior distributions, starting values for all
estimated parameters, processed data matrices, identification
restrictions for structural models, and some parameters determining
particular model features. Apply the \textbf{R6} initialisation by
executing function \texttt{specify\_bsvar\_sv}, using the
package-provided data matrix \texttt{us\_fiscal\_sww}, for a model with
\texttt{p\ =\ 4} lags, as follows:

\begin{verbatim}
spec = specify_bsvar_sv$new(us_fiscal_sww, p = 4)
\end{verbatim}

The created object \texttt{spec} is of class \texttt{BSVARSV} that
includes elements specifying our models that can be accessed as list
elements. Therefore, investigating data matrices can be done by
executing \texttt{spec\$data\_matrices}, which itself is an \textbf{R6}
object of class \texttt{DataMatricesBSVAR} with elements
\texttt{spec\$data\_matrices\$Y} and \texttt{spec\$data\_matrices\$X}.
Similarly, the fixed prior hyper-parameters are collected in element
\texttt{spec\$prior} of class \texttt{PriorBSVARSV}. For instance, the
prior mean of the autoregressive parameters can be accessed via
\texttt{spec\$prior\$A}. A careful modification of this element and
other elements of \texttt{spec\$prior} preserving their type and
dimensions will adjust the prior distributions for the empirical
analysis. Finally, the model identification and starting values can be
accessed and modified via element \texttt{spec\$identification} of class
\texttt{IdentificationBSVARs} and \texttt{spec\$starting\_values} of
class \texttt{StartingValuesBSVARSV}, respectively.

The separation of the model specification and estimation is a
distinguishing feature of our packages that aims at achieving
transparency, flexibility, and possibilities for customisation of the
forecasting workflows. Therefore, the model specification can be
customised by either providing arguments to the initialisation function,
with help available via \texttt{?specify\_bsvar\_sv}, or by modifying
the elements of the created object \texttt{spec}. The specification
object also includes utility functions that are used internally by the
package to manage the workflows, such as \texttt{spec\$get\_normal()} or
\texttt{spec\$get\_identification()}.

\subsection{Forecasting Workflows}\label{ssec:workflow}

\noindent The empirical analysis following the model specification
requires more detailed explanation of the workflow. It is organised
around verbs \texttt{estimate}, \texttt{forecast}, \texttt{compute}, and
\texttt{verify}, and complemented by the \texttt{plot} and
\texttt{summary} functions. These verbs define the generics as well as
the methods for the \textbf{R6} classes, resulting in uniform workflows
for all our packages.

Focusing on the forecasting workflow, consider the first two verbs.
\texttt{estimate} is a generic function defined in package
\textbf{bsvars} with the methods all across our packages specific to the
implemented models. On the other hand, \texttt{forecast} is a generic
imported from package \textbf{generics} by \cite{generics} with the
methods defined for each of our estimation output classes. This solution
allows simultaneous work with other forecasting packages that implement
the same generic function.

We continue the example and estimate the SVAR-SV model. The package
design leads to a two-stage process illustrated by the following code.

\begin{verbatim}
burn = estimate(spec, S = 10000)
post = estimate(burn, S = 10000)
\end{verbatim}

\noindent In the first stage the object \texttt{spec} is provided to
function \texttt{estimate()} with the argument \texttt{S\ =\ 10000} to
obtain 10,000 draws from the posterior distribution via Gibbs sampler.
This step is run to achieve convergence. In fact, this line executes
method \texttt{estimate.BSVARSV} for the class \texttt{BSVARSV} and
returns an object \texttt{burn} of class \texttt{PosteriorBSVARSV} that
includes two elements, namely \texttt{burn\$posterior} with the draws
from the posterior distribution, and \texttt{burn\$last\_draw} of class
\texttt{BSVARSV} whose element
\texttt{burn\$last\_draw\$starting\_values} contains now the last draw
of the MCMC. In the second stage, the object \texttt{burn} is passed to
the method \texttt{estimate.PosteriorBSVARSV}, that initiates the Gibbs
sampler at \texttt{burn\$last\_draw\$starting\_values}, and continues
the MCMC to obtain the final 10,000 draws from the stationary posterior
distribution. The output is saved in object \texttt{post} of class
\texttt{PosteriorBSVARSV} with the same structure as already described.
This object is subsequently used for structural analyses and
forecasting.

The object \texttt{post} containing the estimation output is passed to
function \texttt{forecast()} and complemented by the specification of
the forecast horizon in argument \texttt{horizon}. The following code
calls the method \texttt{forecast.PosteriorBSVARSV} applying the sampler
from the predictive density of the particular model, returns an object
of class \texttt{Forecasts} saved in \texttt{fore}, and plots
predictions using the method \texttt{plot.Forecasts} defined in package
\textbf{bsvars}.

\begin{verbatim}
fore = forecast(post, horizon = 8)
plot(fore)
\end{verbatim}

The \textbf{R6} object of class \texttt{Forecasts} lists the draws from
the predictive density in element \texttt{fore\$forecasts} that can be
used to compute any quantity reporting the forecast, including the mean,
standard deviation, and the 5th and 95th percentiles provided by
executing \texttt{summary(fore)}. The object \texttt{fore} also contains
the mean and covariance of the conditional predictive densities the
draws were generated from in elements \texttt{fore\$forecast\_mean} and
\texttt{fore\$forecast\_covariance}, respectively. They can be used to
compute measures of density forecasting performance, such as the log
predictive score by \cite{ga10}.

Finally, the same workflow can be implemented using the pipe operator
\texttt{\textbar{}\textgreater{}}. As the computational burden is in
estimation, we propose to isolate it in the following code:

\begin{verbatim}
us_fiscal_sww |> 
  specify_bsvar_sv$new(p = 4) |> 
  estimate(S = 10000) |> 
  estimate(S = 10000) -> post
\end{verbatim}

\noindent from other stages of analysis, such as the forecasting
presented in the following line:

\begin{verbatim}
post |> forecast(horizon = 8) |> plot()
\end{verbatim}

\noindent All of these methods are implemented for a wide range of
models collected in our packages granting users access to a variety of
forecasting approaches.

\section{Post-COVID Forecasting of a Monetary-Fiscal System }\label{sec:forecasting}

\noindent The post-COVID forecasting performance of the models from our
developed packages is evaluated using an expanding-window recursive
forecasting exercise for the 10-variable quarterly system proposed by
\cite{mu2009}. The dataset comprises total tax revenue, government
spending, gross domestic product (GDP), the federal funds rate,
consumption, real wages, investment, M2 money stock, the producer price
index of crude materials, and the GDP deflator, covering the period from
Q1 1959 to Q3 2025. The initial forecast origin is Q4 2020. Forecasting
performance is assessed at 1-quarter (\(h=1\)) and 1-year (\(h=4\))
horizons using the predictive log-score (PLS) by \cite{ga10} and the
mean absolute scaled error (MASE) by \cite{hk2006}.

\begin{table}[h]
\begin{center}
\caption{Point and Density Forecasting Results}

\bigskip
\label{tab:fore}
\begin{tabular}{lllllrrrr}
\toprule
 & & & & & \multicolumn{2}{c}{$h=1$} & \multicolumn{2}{c}{$h=4$}\\
id & package & volatility & distr. & order & PLS & MASE & PLS & MASE\\
\midrule
1 & \multirow{11}{*}{\textbf{bsvars}} & ncSV & norm & OI & -13.50 & 0.97 & -18.38 & 1.37 \\
  2 &  & ncSV & norm & LT & -13.08 & 0.95 & -16.99 & 1.37 \\
  3 &  & ncSV & t & OI & -13.61 & 0.97 & -17.91 & 1.37 \\
  4 &  & ncSV & t & LT & -13.26 & 0.96 & -17.11 & 1.37 \\
  5 &  & cSV & norm & OI & -13.43 & 0.96 & -17.72 & 1.39 \\
  6 &  & cSV & norm & LT & -13.21 & 0.95 & -17.11 & 1.39 \\
  7 &  & cSV & t & OI & -13.62 & 0.96 & -17.89 & 1.39 \\
  8 &  & cSV & t & LT & -13.41 & 0.96 & -17.63 & 1.38 \\
  9 &  & const & norm & LT & -15.08 & 1.05 & -18.23 & 1.45 \\
  10 &  & const & t & LT & -14.11 & 0.97 & -17.21 & 1.38 \\
  11 &  & const & t & OI & -14.46 & 0.99 & -18.33 & 1.39 \\
\midrule
  12 & \multirow{6}{*}{\textbf{bvars}} & ncSV & norm & OI & -15.11 & 1.28 & -17.80 & 1.62 \\
  13 &  & ncSV & t & OI & -15.22 & 1.16 & -18.05 & 1.51 \\
  14 &  & cSV & norm & OI & -15.07 & 1.25 & -18.25 & 1.53 \\
  15 &  & cSV & t & OI & -15.32 & 1.15 & -18.15 & 1.51 \\
  16 &  & const & norm & OI & -17.13 & 1.22 & -19.40 & 1.55 \\
  17 &  & const & t & OI & -15.37 & 1.16 & -18.06 & 1.51 \\
\midrule
  18 & \multirow{2}{*}{\textbf{bsvarSIGNs}} & COVID & norm & OI & -16.19 & 1.10 & -18.76 & 1.39 \\
  19 &  & const & norm & OI & -15.94 & 1.10 & -17.80 & 1.31 \\
\bottomrule
\end{tabular}
\end{center}

{\small Note: Data and results can be reproduced following code at: \href{https://github.com/lcq110/A-Design-Concept}{github.com/lcq110/A-Design-Concept}.

Models featured in publications: 
1: \cite{lsuw2026}, 
5: \cite{cky2024},
6: \cite{Clark2011},
8: \cite{chiu2017},
9: \cite{wz2003},
11: \cite{lms2017},
14--17: \cite{chan2020large},
18: \cite{glp2022},
19: \cite{glp2015},
2--4,7,12,13: this paper.
}
\end{table}

Our model performance assessment focuses on comparing models with
centred SV exhibiting excellent performance according to
\cite{Clark2011}, \cite{carriero2016common}, \cite{chan2020large}, and
\cite{cky2024}, to those with non-centred SV proposed by
\cite{lsuw2026}. Overall, the comparison includes 19 models, which
differ according to the features presented in Table~\ref{tab:fore}.
These features include volatility processes: non-centred SV (ncSV),
centred SV (cSV), and the homoskedastic model (const)---as well as error
term distributions, which are set to either normal (norm) or Student-t
(t). Additionally, 11 structural models from the package \textbf{bsvars}
differ in terms of order invariance, which is present in models with an
unrestricted matrix \(\mathbf{B}\) (OI) and absent in models with a
lower-triangular \(\mathbf{B}\) (LT). Forecasts are also generated using
reduced-form models, including 6 from the package \textbf{bvars} and 2
from \textbf{bsvarSIGNs}. All eight reduced-form models are order
invariant, as their covariance is estimated explicitly by the matrix
\(\boldsymbol\Sigma\) and they feature common volatility. These models
differ along features similar to those of the structural models, except
for model 18, which incorporates COVID-specific volatility by
\cite{glp2022}. Many of these models were used in prominent studies
listed in the note to Table~\ref{tab:fore}.

We summarise the forecasting performance by reporting average predictive
scores, calculated as the mean exponent of the PLS for each feature at
each forecast horizon. At both forecast horizons, non-centred SV models
outperform centred SV models. In particular, non-centred SV raises the
average exponentiated PLS by approximately 6\% at \(h=1\) and 15\% at
\(h=4\) relative to centred SV, with similar tendencies holding when
structural or reduced-form models are considered only. Analysing the
average MASE, alternative SV parameterisations yield very similar point
forecast accuracy, with relative differences below 1\%. Additionally,
according to both performance measures t-distributed models outperform
normal models and structural models outperform reduced-form models at
both horizons.

\section{Conclusions}

\noindent We present a suite of open-source packages designed to provide
a unified framework for Bayesian VAR forecasting. Their novel
combination of features makes scalable and reproducible predictions
possible using normalised heteroskedastic non-Gaussian models. The
reported forecasting performance measures show that these features are
essential to obtain precise forecasts, which emphasises the importance
of our software design.

\noindent  \setstretch{1.5}

\bibliography{bsvarsFORE.bib}

\end{document}